\documentclass[iop]{emulateapj}

\shorttitle{Discovery of SiO band emission from Galactic B[e] supergiants}
\shortauthors{Kraus et al.}

\begin{document}


\title{Discovery of SiO band emission from Galactic B[e] supergiants $^*$}

\altaffiltext{*}{Based on
observations collected with the ESO VLT Paranal Observatory under program 
093.D-0248(A).}

\author{M. Kraus}
\affil{Astronomick\'y \'ustav, Akademie v\v{e}d \v{C}esk\'e republiky,  
Fri\v{c}ova 298, 251\,65 Ond\v{r}ejov, Czech Republic}
\email{michaela.kraus@asu.cas.cz}

\author{M. E. Oksala}
\affil{LESIA, Observatoire de Paris, CNRS UMR 8109, UPMC, Universit\'{e} Paris Diderot, 5 place Jules Janssen, 92190, Meudon, France}

\author{L. S. Cidale, M. L. Arias, A. F. Torres}
\affil{Departamento de Espectroscop\'ia Estelar, Facultad de Ciencias
Astron\'omicas y Geof\'isicas, Universidad Nacional de La Plata \\ Instituto 
de Astrof\'isica de La Plata, CCT La Plata, CONICET-UNLP, Paseo del
Bosque s/n, B1900FWA, La Plata, Argentina}
 
\and

\author{M. Borges Fernandes}
\affil{Observat\'orio Nacional, Rua General Jos\'e Cristino 77,
20921-400 S\~ao Cristov\~ao, Rio de Janeiro, Brazil}

\begin{abstract}
B[e] supergiants (B[e]SGs) are evolved massive stars in a
short-lived transition phase. During this phase, these objects eject large 
amounts
of material, which accumulates in a circumstellar disk-like structure.
The expelled material is
typically dense and cool, providing the cradle for molecule and dust 
condensation and for a rich, ongoing chemistry. Very little is known about the chemical composition of these disks, beyond the 
emission from dust and CO revolving around the star on Keplerian orbits.
As massive stars preserve an oxygen-rich surface composition throughout their
life, other oxygen-based molecules can be expected to form.
As SiO is the second most stable oxygen compound, we initiated an observing 
campaign to search for first-overtone SiO emission bands. We obtained 
high-resolution near-infrared $L$-band spectra for a sample of Galactic 
B[e]SGs with reported CO band emission. We clearly detect emission 
from the SiO first-overtone bands in CPD-52\,9243 and indications for  
faint emission in HD\,62\,623, HD\,327\,083, and CPD-57\,2874. From model fits, 
we find that in all these stars the SiO bands are rotationally broadened 
with a velocity lower than observed in the CO band forming regions, suggesting 
that SiO forms at larger distances from the star. Hence, searching for and 
analyzing these bands is crucial for studying the structure and kinematics 
of circumstellar disks, because they trace 
complementary regions to the CO band formation zone. Moreover, since SiO
molecules are the building blocks for silicate dust, their study might provide 
insight in the early stage of dust formation.  
\end{abstract}


\section{Introduction}

Massive stars, when evolving off the main sequence, undergo phases of strong 
mass-loss, often resulting in the formation of shells, disks or rings of 
circumstellar material. One class of evolved massive stars, B[e] supergiants 
(B[e]SGs), are early-type emission line stars with high-density gaseous and 
dusty circumstellar disks of yet unknown origin \citep[see][for a recent 
review]{2014AdAst2014E..10D}. The gaseous inner disk is 
traced by the emission of low-ionized metal lines from both permitted and
forbidden transitions \citep[e.g.,][]{1985A&A...143..421Z}. Of these, the 
lines of 
[Ca\,II] and [O\,I] 
are particularly useful. While their 
intensities probe regions of high and medium densities,
where the transition between the two regimes might be settled
at an electron density of $\sim 10^{7}$\,cm$^{-3}$, their line profiles 
contain the information on their kinematics \citep{2007A&A...463..627K, 
2010A&A...517A..30K, 2012MNRAS.423..284A}. Farther out, molecules form via gas 
phase chemistry, and intense band emission from CO molecules has been reported
for many of these stars \citep{1988ApJ...324.1071M,
1988ApJ...334..639M, 1989A&A...223..237M, 1996ApJ...470..597M,
2000A&A...362..158K, 2014ApJ...780L..10K, 2010MNRAS.408L...6L, 
2012A&A...548A..72C, 2012A&A...543A..77W, 2012MNRAS.426L..56O, 
2013A&A...558A..17O}. In addition, the presence of TiO band emission features 
in optical spectra of several B[e]SGs has been suggested 
\citep{1989A&A...220..206Z, 2012MNRAS.427L..80T}, but still lacks confirmation.
Besides atomic and molecular gas, these stars are also surrounded by warm 
circumstellar dust, as is obvious from their strong infrared excess emission 
\citep{1986A&A...163..119Z, 2009AJ....138.1003B, 2010AJ....140..416B}, 
interferometric observations \citep{2007A&A...464...81D, 2011A&A...526A.107M,
2012A&A...548A..72C}, and resolved spectral dust features 
\citep{2006ApJ...638L..29K, 2010AJ....139.1993K}.

Kinematic studies of the atomic and CO gas reveal that the material 
is confined to detached (sometimes multiple) rings rotating around the star on 
Keplerian orbits \citep[e.g.,][]{2010A&A...517A..30K, 2013A&A...549A..28K, 
2014ApJ...780L..10K, 2012MNRAS.423..284A, 2012A&A...543A..77W, 
2012A&A...548A..72C, 2014arXiv1409.7550M}. Abundance studies of these rings 
reveal an enrichment in the isotopic molecule $^{13}$CO, in agreement with an 
evolutionary stage just beyond the main sequence \citep{2009A&A...494..253K, 
2010MNRAS.408L...6L, 2013A&A...549A..28K, 2013A&A...558A..17O, 
2014arXiv1409.7550M}. 

Furthermore, 
B[e]SGs could be the progenitors of a group of early-type blue supergiants 
(BSGs) with ring nebulae such as, e.g., Sher\,25 and SBW\,1. Optical images of 
these two stars display equatorial rings and bipolar lobes, and 
\citet{2007AJ....134..846S} propose that as soon as the rings or disk-like 
structures seen around B[e]SGs have expanded and cooled they would 
look exactly like these objects.
Support for such a possible evolutionary link comes from the fact that,
like B[e]SGs, neither Sher\,25 nor SBW\,1 has evolved yet through a red
supergiant phase, indicating that in both cases, the ejection of the material
forming the disks or rings must have happened during their BSG phase.
Such a scenario has a much wider impact: as the BSGs with rings
closely resemble the progenitor star of the supernova SN 1987A, 
B[e]SGs might be regarded as the most plausible candidates
for supernova explosions of the SN 1987A-type.
Notably, two B[e]SGs were recently suggested to be viable SN\,1987A-type 
progenitor candidates: \object{LHA\,115-S\,18} in the Small Magellanic Cloud 
\citep{2013A&A...560A..10C} and the Galactic object MWC\,137
\citep{2014arXiv1409.7550M}.
 For a better understanding of
the evolution of these stars up to the supernova stage, it is hence essential
to study B[e]SGs in great detail, and in particular their mass-loss behavior 
and the chemistry involved in the formation of the observed dense molecular and 
dusty rings or disk-like structures.

The atmospheres of massive stars have an oxygen-rich composition, which means 
that the abundance of oxygen atoms greatly exceeds that of carbon atoms. 
The same trend is then expected in their disks, because they form from the
material released from the stellar surface. Consequently, carbon atoms in the 
disks are locked in CO, which, as the most stable molecule with a 
dissociation energy of 11.2\,eV, forms as soon as the gas temperature drops 
below the dissociation temperature of CO, for which a value of 
$\sim 5000$\,K is typically quoted. The excess oxygen atoms will form other 
molecules, and a rich chemistry is expected at larger distances from the star.

The SiO molecule is the second most stable oxygen compound. Its binding energy 
of 8.0\,eV is higher than that of other oxides or oxygen compounds such 
as NO, OH, or water, but lower than that of CO, and thus SiO should form in 
slightly cooler environments ($T_{\rm SiO} < T_{\rm CO} < 5000$\,K). 
As with CO, the rather high gas temperature of the SiO forming region
guarantees that high vibrational levels will be excited, causing coupled
rotational-vibrational transitions, detectable via prominent band and band head 
features. And in fact, SiO bands are commonly 
observed in absorption in the atmospheres of Mira stars, cool giants,
and red supergiants
\citep{1976ApJ...210L.141H, 1979ApJ...230L..47G, 1997A&A...323..202A,
1999A&A...342..799A}. In these objects, SiO is an important molecule, based on
which the physical properties in the transition region between the outer
atmosphere and the innermost circumstellar envelope can be studied
\citep[e.g.,][]{2014A&A...561A..47O}.

To our knowledge, rotational-vibrational SiO bands in emission have only 
been reported for
Supernova 1987A \citep{1988MNRAS.235P..19A, 1991MNRAS.252P..39R, 
1994ApJ...428..769L}. These emission bands appeared 160 days 
after the outburst and were not detectable beyond day 578.
The rise of the SiO features had a time-delay of 48 days with respect to 
that of CO molecular bands. This is in agreement with the lower binding 
energy, requiring a cooler environment for the formation of SiO molecules
by gas phase chemistry. The disappearance of the emission bands agrees 
with the onset of dust formation in the ejecta, which is interpreted as
depletion of molecular SiO due to condensation of silicate dust, for which
the SiO molecule is proposed to be a major building block within
an oxygen-rich environment. 

Unlike the transient phenomena of a rapidly expanding, diluting ejecta of a 
supernova, the rings or disks around B[e]SGs seem to be quite stable 
structures of accumulated material. This guarantees these disks
as ideal chemical laboratories to study molecule formation and
dust condensation. However, apart from CO band
emission, detected in most of these objects, and warm circumstellar dust,
no information exists about the chemical composition of these disks.
Hence, it is necessary to begin searching for other molecular emission features 
that can be used to study the structure and kinematics of B[e]SG stars' disks
comprehensively.

To pioneer this work, we selected a sample of Galactic B[e]SGs, with
known CO band emission, to search for molecular rotational-vibrational 
band emission from SiO.

\section{Observations}

During the period of March-May 2014, we observed four Galactic B[e]SGs in the 
$L$-band wavelength region using the CRyogenic high-resolution InfraRed Echelle 
Spectrograph \citep[CRIRES,][]{2004SPIE.5492.1218K} on the ESO VLT UT1-Antu 
8-m telescope. Details on the objects and their observations are given in 
Table\,\ref{tbl_obs}. Data were obtained at four different standard 
settings with reference wavelengths of 4048.6, 4060.1, 4135.5, and 4147.0 nm, 
so that the total spectra cover a continuous wavelength range from 
3.988--4.176\,$\mu$m. This range contains the first four band heads of the SiO
first-overtone bands. We used the 0.4\arcsec slit, which provides a spectral 
resolution of $R\sim 50\,000$, corresponding to a velocity resolution of
$\sim 6$\,km\,s$^{-1}$. 

Data reduction was performed with the ESO CRIRES pipeline (version 2.3.2).
The observations were taken in an ABBA nod pattern for proper sky subtractions.
Raw frames were corrected for bad pixels, flat fields, non-linearity, and then 
wavelength calibrated using OH lines in the standard star spectrum.
A telluric standard star observation was similarly reduced. For telluric 
correction, a B-type standard star was observed immediately after the target at 
a similar airmass. The IRAF\footnote{IRAF is distributed by the National Optical
Astronomy Observatory, which is operated by the Association of Universities for
Research in Astronomy (AURA) under cooperative agreement with the National
Science Foundation.} task \textit{telluric} was used to remove the atmospheric 
features. The final spectra were normalized and continuum subtracted.

\section{Results}

\begin{figure}
\epsscale{0.80}
\plotone{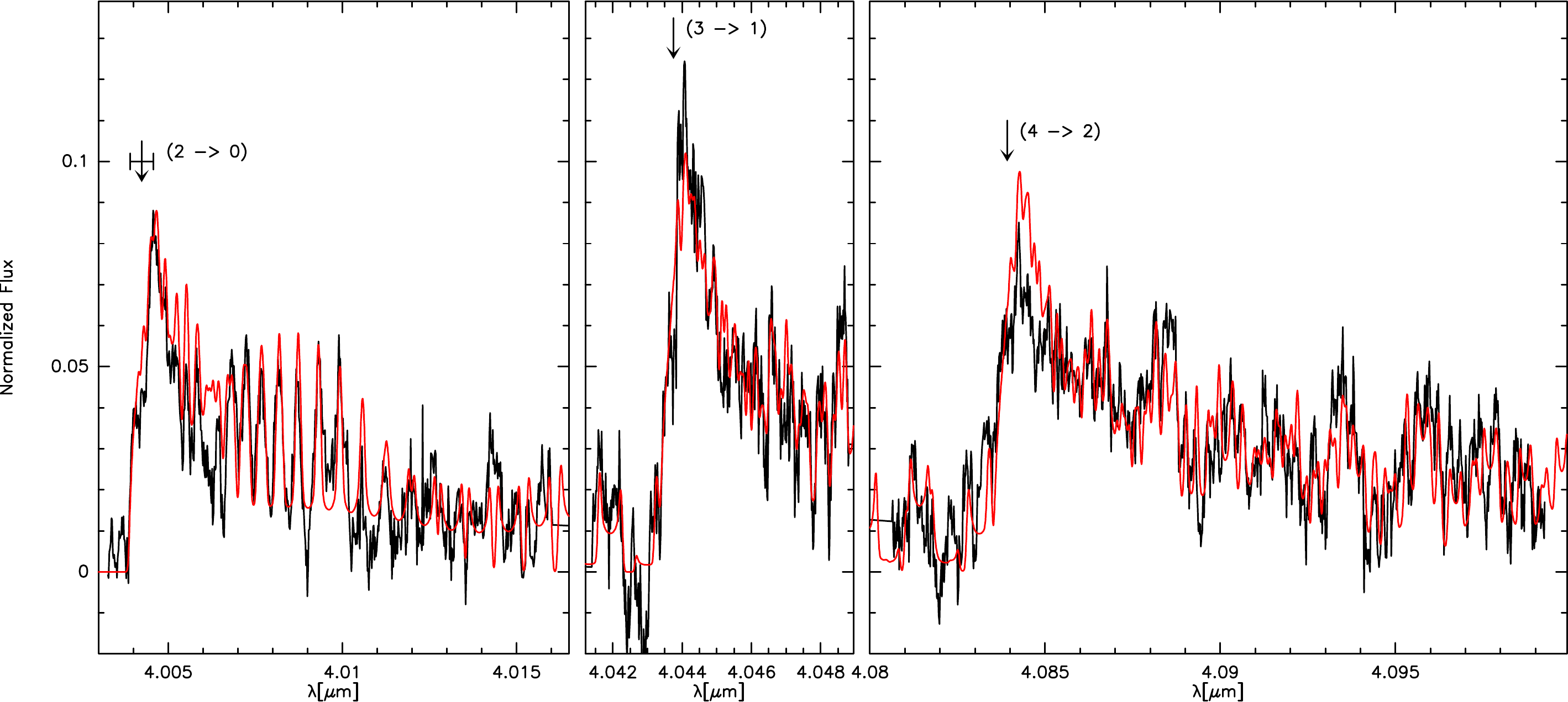}
\caption{Model fit (red) to the observed (black) three band heads of the SiO
first-overtone band emission detected in CPD-52\,9243. The arrows mark
the wavelength of the unbroadened band heads, and the bar in the panel of the first band head marks the separation between the blue-shifted shoulder and the red-shifted maximum.}
\label{fig_cpd-fit}
\end{figure}

\begin{figure}
\epsscale{0.50}
\plotone{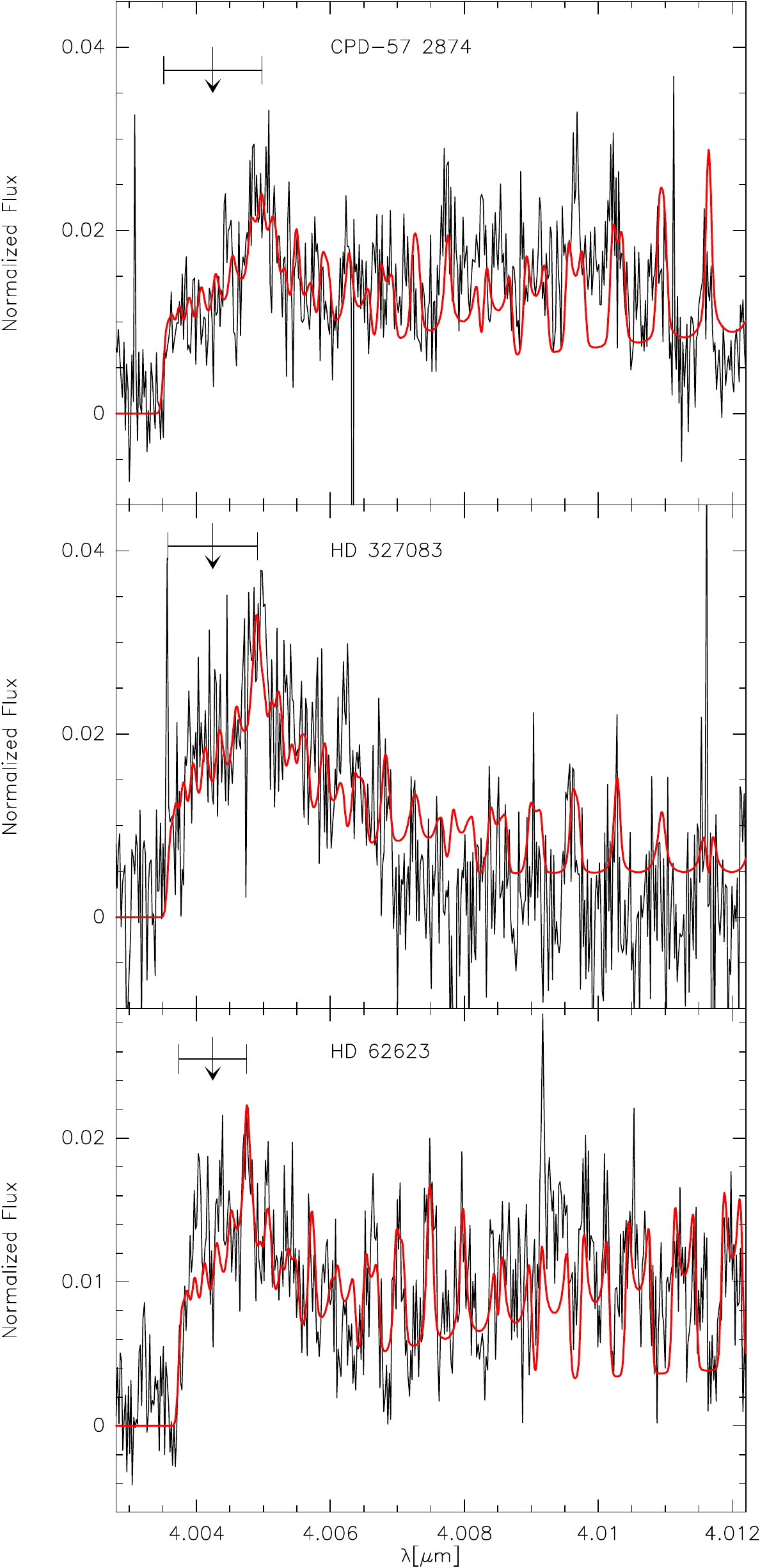}
\caption{Model fits (red) to the observed (black) first SiO band head. The
arrow marks the wavelength of the unbroadened band head, and the bar marks
the  separation between the blue-shifted shoulder and the red-shifted maximum.}
\label{fig_sio}
\end{figure}

We clearly detect emission from the first three band heads of the SiO 
first-overtone bands in CPD-52\,9243 (Fig.\,\ref{fig_cpd-fit}), and from
the first band head in the spectra of the other three stars, CPD-57 2874, 
HD\,327083, and HD\,62623 (Fig.\,\ref{fig_sio}). 

In each star's spectrum, the band heads display a blue-shifted shoulder and a 
red-shifted maximum with respect to the laboratory wavelength. Their
separation is marked by the bar in Figs.\,\ref{fig_cpd-fit} and \ref{fig_sio}. 
Such structures are also typically observed in the band heads of CO, and are an 
unequivocal manifestation of rotational broadening.

To confirm that the observed features are emission from
SiO molecules, we compute synthetic band head models. For this, we modify our 
existing CO disk code \citep{2000A&A...362..158K} by implementing the SiO line 
lists and Einstein transition probabilities provided by \citet{Barton2013}. For 
each of our targets, information about the rotation velocity of the CO region 
and disk inclination angles are listed in Table\,\ref{tbl_param}. The 
rotation velocities of the CO gas have been obtained from the broadening of the 
first band head of the CO first-overtone bands resolved in high-resolution 
spectra, and the disk inclination angles have been determined from 
interferometric observations. From the wavelength separation between the 
position of the blue shoulder and the red peak of the first band head, we 
obtain an estimate of the SiO rotation velocity projected to the line of 
sight. This parameter is refined during the fitting procedure. The final 
rotation velocities, corrected for disk inclination, are included in 
Table\,\ref{tbl_param}. The high spectral resolution, combined with the 
sensitivity of the synthetic spectra to rotation velocity, guarantee that these 
values have a precision of $\pm 1$\,km\,s$^{-1}$. 
Comparing the velocities obtained in the two molecular band emitting regions, 
we find a consistently lower value for SiO, suggesting that this molecule is 
formed within the Keplerian disk at larger distances from the star.

For CPD-52 9243, where three observed band heads are available, we can also 
constrain temperature and column density, for which we find $T_{\rm SiO} = 
2000\pm 200$\,K and $N_{\rm SiO} = (5\pm 2)\times 10^{21}$\,cm\,$^{-2}$.
With this column density, most of the individual SiO rotation-vibration 
lines within the observed wavelength range are optically thick.
Although the SiO rotation velocity in this object is only marginally lower, 
both temperature and column density are significantly lower than the values 
obtained from the CO modeling \citep[see][]{2012A&A...548A..72C}. This 
suggests that the two band head formation regions are physically disjoint. 

For the other three objects, where solely the first band head is detected,
it is only possible to constrain the 
rotation velocity, because different combinations of temperature and density
result in almost identical synthetic spectra. Hence, the fits shown in 
Fig.\,\ref{fig_sio} are only to demonstrate that the observed features can
indeed be assigned to SiO band emission, and to determine the SiO rotation 
velocity.

Under the assumption of Keplerian rotation of the disks and using the 
stellar mass ranges of the objects (Table\,\ref{tbl_obs}), the radii of both 
(CO and SiO) molecular rings can be computed. These values are included in
Table\,\ref{tbl_param}. Furthermore, considering that the circumstellar dust
is mainly composed of silicates, we can calculate the dust evaporation 
distances. Assuming that the grains are spheres with a maximum size of $\sim
1\,\mu$m, and that the silicate evaporation temperature is $\sim 1500$\,K, the 
minimum dust evaporation radii follow from the stellar luminosities (listed
in Table\,\ref{tbl_obs}) and are added to Table\,\ref{tbl_param}. Obviously,
the much hotter molecular rings reside within the dusty disks.

\section{Conclusions}

We report on the detection of SiO first-overtone band emission in the
high-resolution $L$-band spectra of four Galactic B[e]SGs, which are known to 
display strong CO band emission. While SiO bands are commonly seen in 
absorption from the atmospheres of cool stars and giants, they were detected in 
emission, so far, only in spectra of SN\,1987\,A during the post-explosion
period from day 160 to day 578.
Hence, to our knowledge, this is the first discovery of SiO 
first-overtone band emission from hot, high-density environments of
evolved massive stars.

The high spectral resolution of our data allows us to precisely determine 
the rotation velocity of the SiO gas. For all of our target stars, this 
velocity is lower than that determined from the CO band forming region. 
Moreover, model fits to the observed three band heads in CPD-52 9243 
demonstrate that, in contrast to the CO band emitting region, the emission of 
the SiO bands originates from a cooler and less dense disk region, indicating 
that SiO molecules form at larger distances within the Keplerian disks. 
For all four objects, we find, in addition, that the SiO ring radii are smaller 
than the silicate dust evaporation distances.

This finding is of great importance, because it demonstrates that molecules
provide an excellent tool to study B[e]SG stars' disks. The emission features 
of both molecules detected so far, CO and SiO, carry all the essential 
information about the physical properties (temperature, density) and kinematics 
of their formation region. As different molecules require diverse conditions, a 
huge variety can be expected to form throughout the disk, filling the space
between the hottest inner rim traced by CO bands, and the dust condensation
region. Hence, it is essential and timely to search for emission features from 
many more molecules that are expected to form within the oxygen-rich 
environment of these disks. Moreover, molecules such as CO and SiO are 
also indispensable to study the disk structure at much larger distances. With
their numerous pure rotational transitions, these molecules are also suitable
to trace much cooler environments. The rotational transitions spread over the 
submm and radio regime, making the disks of B[e]SGs ideal targets to be 
observed with high spatial and spectral resolution facilities such as ALMA.
Combining the results obtained from different 
molecular species tracing diverse temperature regions, and hence distances from 
the star, will greatly improve our knowledge and comprehension of the disk 
structure, and help to understand the disks' possible formation history. 

In addition, the discovery of SiO band emission has an even wider impact. 
Evolved massive stars, such as red supergiants, B[e]SGs, and supernovae 
(e.g., SN\,1987A), are the main sources of silicate dust, for which SiO 
molecules are 
regarded as a major building block. Studying the physical properties of SiO gas 
in the disks of B[e]SGs might, hence, provide indispensable insight into the 
early stage of silicate dust formation within the oxygen-rich environment of 
these enigmatic stars.

\acknowledgments

We thank the referee, Dr. John Bally, for his valuable comments on the 
manuscript. This research made use of the NASA Astrophysics Data System (ADS) 
and of the SIMBAD database, operated at CDS, Strasbourg, France. 
MK acknowledges financial support from GA\v{C}R (grant number 14-21373S). The
Astronomical Institute Ond\v{r}ejov is supported by the project RVO:67985815.
LC, MLA, and AFT acknowledge financial support from the Agencia de Promoci\'on 
Cient\'{\i}fica  y Tecnol\'ogica (Pr\'estamo BID PICT 2011/0885), CONICET 
(PIP 0300), and the Universidad Nacional de La Plata (Programa de Incentivos 
G11/109), Argentina. Financial support for International Cooperation of the 
Czech Republic (M\v{S}MT, 7AMB14AR017) and Argentina (Mincyt-Meys ARC/13/12 and 
CONICET 14/003) is acknowledged.

\clearpage

\begin{deluxetable}{lccccccccr}
\tabletypesize{\scriptsize}
\tablecaption{Observed objects\label{tbl_obs}}
\tablewidth{0pt}
\tablehead{
\colhead{Object}  & \colhead{$\alpha$ (J2000)} & \colhead{$\delta$ (J2000)} &
\colhead{K} & \colhead{Obs. Date} & \colhead{$T_{\rm{eff}}$} &
\colhead{$M_\star$} &  \colhead{$\log\,L/L_\odot$} &\colhead{$D$} &  
\colhead{Ref.}\\
                  &                            &                            &
 (mag)      &                     &     (K)                  &
  ($M_\odot$)       &                              & (kpc)        &  }
\startdata
HD 62623    & 07 43 48.469 & -28 57 17.37 & 2.340 & 2014-03-27 & 8\,500--9\,500 & 31--39 & 4.8--5.2 & 1.7$\pm$0.5 &  1, 2 \\
CPD-57 2874 & 10 15 21.971 & -57 51 42.71 & 4.281 & 2014-04-16 & 9\,000--10\,000  & 15--20 & 6.00 & 1.7$\pm$0.7 & 3\\
CPD-52 9243 & 16 07 01.968 & -53 03 45.75 & 4.443 & 2014-03-22 & 15\,800  & 17.4--18.6 & 4.97 & 3.44$\pm$0.8 & 4, 5\\
HD 327083   & 17 15 15.374 & -40 20 06.74 & 3.296 & 2014-05-31 & 11\,500 & 60 & 6.00 & 4.80 & 6 \\
            &              &              &       &            & 20\,000 & 25 & 5.00 & 1.50$\pm$0.5 & 7
\enddata
\tablerefs{(1) \citet{1995A&A...293..363P}; (2) \citet{2007A&A...474..653V}; 
(3) \citet{2011A&A...525A..22D}; (4) \citet{2012A&A...548A..72C}; 
(5) \citet{1981A&A....98..112S}; (6) \citet{2003A&A...409..665M}; 
(7) \citet{2003A&A...406..673M}.}
\end{deluxetable}


\begin{deluxetable}{lccccccc}
\tabletypesize{\scriptsize}
\tablecaption{Rotation velocities and radii of the CO and SiO disk regions, and silicate dust evaporation distances \label{tbl_param}}
\tablewidth{0pt}
\tablehead{
\colhead{Object} & \colhead{$i$} & \colhead{$v_{\rm rot}$(CO)} &
\colhead{References} & \colhead{$v_{\rm rot}$(SiO)} & \colhead{$r$(CO)} & 
\colhead{$r$(SiO)} & \colhead{$r_{\rm evap}$(silicates)} \\
 & ($\degr$) & (km\,s$^{-1}$) & & (km\,s$^{-1}$) & (AU) & (AU) & (AU)}
\startdata
CPD-52 9243 & 46 &  36 & 1    & 35.5 & 11.9--12.7 & 12.2--13.1 & 16.1 \\
CPD-57 2874 & 60 & 130 & 2, 3 & 110  &  0.8--1.0  &  1.1--1.5  & 52.7 \\
HD 327083   & 50 &  86 & 4    & 78   &  3.0 / 7.2 &  3.6 / 8.7 & 16.7 / 52.7\\
HD 62623    & 38 &  53 & 3, 5 & 48   &  9.8--12.3 & 11.9--15.0 & 17.5
\enddata
\tablerefs{(1) \citet{2012A&A...548A..72C}; (2)
\citet{2011A&A...525A..22D}; (3) \citet{2012ASPC..464...67M}; (4)
Andruchow et al. (in preparation); (5) \citet{2011A&A...526A.107M}.}
\end{deluxetable}

\end{document}